\documentclass[%
 preprint,showpacs,preprintnumbers,amsmath,amssymb,aps]{revtex4-1}
\usepackage{graphicx}
\usepackage{dcolumn}
\usepackage{bm}
\usepackage{amssymb}
\usepackage{multirow}
\usepackage{bigstrut}
\usepackage{makecell,rotating}

\usepackage{mathrsfs}
\usepackage{booktabs}
\usepackage{threeparttable}
\usepackage{multirow}
\usepackage{epsfig}
\usepackage{threeparttable}
\usepackage{chngpage}
\usepackage{float}

\usepackage{amstext}
\usepackage{amsmath}

\usepackage{lineno}

\newcommand{\X}{\mathbf{x}}
\newcommand{\A}{\mathbf{A}}
\newcommand{\B}{\mathbf{B}}
\newcommand{\D}{\mathbf{d}}
\newcommand{\W}{\mathbf{W}}
\newcommand{\U}{\mathbf{u}}

\newcommand{\zero}{\mathbf{0}}

\newcommand{\ie}{\emph{i.e.}}

\usepackage[top=23.4mm, bottom=23.4mm, left=23.4mm, right=23.4mm]{geometry}

\begin{document}

\title{Controlling network dynamics}
\author{Aming Li$^{1, 2,}$\footnote{aming.li@zoo.ox.ac.uk}, and Yang-Yu Liu$^{3, 4,}$\footnote{spyli@channing.harvard.edu}}
\affiliation{
\begin{enumerate}
  \item Department of Zoology, University of Oxford, Oxford OX1 3PS, UK
  \item Department of Biochemistry, University of Oxford, Oxford OX1 3QU, UK
  \item Channing Division of Network Medicine,
Brigham and Women's Hospital
and Harvard Medical School, Boston, Massachusetts 02115, USA
  \item Dana-Farber Cancer Institute, Boston, Massachusetts 02115, USA
\end{enumerate}
}
\date{\today}

\begin{abstract}
Network science has experienced unprecedented rapid development in the past two decades. The network perspective has also been widely applied to explore various complex systems in great depth.
In the first decade, fundamental characteristics of complex network structure, such as the small-worldness, scale-freeness, and modularity, of various complex networked systems were harvested from analyzing big empirical data.
The associated dynamical processes on complex networks were also heavily studied.
In the second decade, 
more attention was devoted to investigating the control of complex networked systems, ranging from fundamental theories to practical applications.
Here we briefly review recent progress regarding network dynamics and control, mainly concentrating on research questions proposed in the six papers we collected for the topical issue entitled ``Network Dynamics and Control'' at \textit{Advances in Complex Systems}.
This review closes with possible research directions along this line, and several important problems to be solved.
We expect that,  in the near future, network control will play an even bigger role in more fields, helping us understand and control many complex natural and engineered systems.
\end{abstract}


\maketitle

\section{Introduction}
The world is flooded with various complex networked systems. 
Typical examples range from the man-made jungle of the interlinked World Wide Web (WWW) documents that billions of people use to interact over the Internet, to highly sophisticated biochemical or biophysical interactions among proteins, DNA, RNA and small molecules in precise cellular processes where even a tiny aberration might lead to a disease.
Starting from capturing the underlying skeleton, networks or graphs provide a common ground to explore these systems, where nodes (a.k.a. vertices, \textit{e.g.}, documents or other web resources in the network of WWW) represent system components and links (a.k.a. edges, \textit{e.g.}, hyperlinks in the network of WWW) indicate their interactions \cite{Havlin2004book,Yangreview15}.
For example, scale-free networks \cite{Barabasi1999a} discovered twenty years ago suggest that, over the whole network, the probability distribution of the number of links each node has (referred to as node's degree) generally follows a power-law distribution for many real-world systems.
This holds true for systems like the network of WWW documents connected by hyperlinks \cite{Barabasi1999b}, and metabolic networks where metabolites  are linked together by chemical reactions \cite{Jeong2000}. 

The power-law degree distribution and a few other network characteristics such as high clustering and modularity capture key topological features of affiliated networked systems. 
Indeed, components in complex natural or engineered systems are often not connected randomly, which further ends in the Poisson distribution quantitatively \cite{Erdos1960}.
Rather, they form heterogeneous networks
with clustered cliques \cite{Watts1998a} and communities \cite{Girvan02PNASCommunity},
where a few hubs are connected with a large number of low-degree nodes \cite{Barabasi1999a}.
Apart from the systems' architectural information, the scale-free nature of real networks also conveys the systems' impressive robustness to random failure of nodes, and fragility to targeted attacks on hubs \cite{AlbertAttack2000,CohenAttack}.
Moreover, for dynamical processes taking place in these systems, emergent macroscopical phenomena such as epidemics spreading \cite{Pastor2001aEpideNoThers} and  cooperative behavior \cite{Santos05PRL} are frequently observed.
In the past two decades, complex networks research experiences an explosion as it keeps bringing scientists from, for example, physics, biology, applied mathematics, social and computer science, etc., to explore various complex systems.
This unparalleled trend eventually facilitates the emergence of an independent area -- network science.

\section{Network controllability}
The rapid development of network science on understanding the structure and dynamics of complex systems prompts us to consider the controllability of these systems -- an important goal to study complex systems. 
Indeed, control is essential for most real systems \cite{Kalman63,Guo1994q,SimonLevin1999,ConRealSysCell11,ConRealSysPNAS12}.
The investigations on the controllability of complex networks are not only because control is a proof of our accumulated deep understanding of many complex systems, but also because control can offer us an operable way to steer the complex system towards any desired state.
Precisely, according to classical control theory, a system is defined as controllable if it can be steered from any initial state to any final state in finite time with admissible control inputs \cite{Kalman63,ReinschkeKJbook,ChenChiTsongbook,Trentelmanbook2012}.

Mathematically, the time evolution of the system states could be modelled by different dynamics \cite{Kalman63,Cornelius2013NatCommun,Wanglezhi16NC,Motter15PRX,ACSAlbert,Duan2019ACS,ACSBabak,ACSXiang,ACSCao,Frank2019,Li2017Sci,Wangling2007,Guo2003,Chendaizhan2009,Wanglong2010,Wanglong2004,Guanyongqiang2013,Xie2002tac}.
For simplicity and understanding the general framework of network control intuitively, here we introduce 
the classical continuous linear time-invariant dynamics
\begin{equation} 
\label{EquDynNode}
\frac{\text{d} \X (t)}{\text{d} t}  = \A \X(t) + \B\U(t), 
\end{equation}
where $\X(t)=(x_1(t),\cdots,x_N(t))^{\text{T}}$ captures the system state at time $t$ with $N$ nodes, and a single node $i$'s state is $x_i(t)$ accordingly.
For example, in the context of brain networks, the state $x_i(t)$ can be regarded as state of neurophysiological activity of the $i$th brain region \cite{Gu2016,TangReviewRMP,Bassett18NP}. 
The square matrix $\A=(a_{ij})_{NN}$ with size $N$ encodes the network structure of the system, where $a_{ij}=0$ if there is no link between nodes $i$ and $j$, otherwise the nonzero $a_{ij}$ represents how, and at which strength, node $j$ affects $i$.
External inputs are depicted by the vector $\U(t) = (u_1(t),\cdots,u_p(t))^{\text{T}}$ with $p \leq N$. 
The input matrix $\B$ with entries only $0$ and $1$ of size $N \times p $, defines that the node $i$ is controlled by the input $u_q(t)$ if $b_{iq} = 1$ ($q \leq p$).

By solving the above equation, we have the system state 
\begin{equation*}
\X(t_\text{f}) = \mathrm{e}^{\A (t_\text{f} - t_0)} \X_0+ \int_{t_0}^{t_\text{f}} \mathrm{e}^{\A (t_\text{f} - \tau)} \B \U(\tau) \mathrm{d} \tau
\end{equation*}
at $t_\text{f}$. 
Based on the definition of network controllability, when the system is controllable, there must exist a control input $\U(t)$ with $t\in[t_0, t_\text{f}]$ to actuate the system state from $\X_0$ at $t_0$ to  $\X_\text{f}$ at $t_\text{f}$, for arbitrary values of $\X_0$, $t_0$, $\X_\text{f}$, and $t_\text{f}$. 
The examination regarding the existence of control inputs could be performed by the Kalman's rank condition \cite{Kalman63}, which states that the system (\ref{EquDynNode}) is controllable if and only if the controllability matrix $\mathbf{C(\mathbf{A}, \mathbf{B})} = [ \mathbf{B},  \mathbf{AB}, \cdots, \mathbf{A}^{N-1}\mathbf{B}]$ has full row rank.

Nevertheless, the rank of $\mathbf{C}(\mathbf{A}, \mathbf{B})$ strongly relies on all the exact entries of both $\A$ and $\B$.
For systems where the existence of links is known (namely, we only know that some entries are zero) but the precise parameters (non-zero entries) are hard to recognize, we could resort to the concept of structural controllability \cite{Lin1970}. 
The system (\ref{EquDynNode}) is structurally controllable if appropriate parameters could be found for non-zero entries in both $\A$ and $\B$ such that $\mathbf{C}(\mathbf{A}, \mathbf{B})$ has full rank accordingly.
As a function of enormous configurations of non-zero entries in $\A$ and $\B$, the rank of $\mathbf{C}(\mathbf{A}, \mathbf{B})$ gets its maximum value (also known as the generic rank \cite{JOHNSTON:1984aa}) for almost all feasible configurations except for some pathological scenarios with Lebesgue measure zero \cite{Lin1970}.
Therefore, structural controllability offers a concrete way to explore how the network structure itself affects our ability to control the corresponding system \cite{Liu2011,Ruths2014science,TangReviewRMP}.

\section{Driver nodes}

To investigate the structural controllability of complex networks, Liu \textit{et al.} \cite{Liu2011} proposed an efficient algorithm to determine the minimal driver node set (MDNS) in which all nodes need to be driven by independent external control inputs for arbitrary directed networks (see Fig.~\ref{Fig_nodes}). 
They found that the size of the MDNS is mainly determined by the underlying degree distribution for many different types of real networks.
Specifically, sparse and heterogeneous networks turn out to be the most difficult to control compared to dense and homogeneous networks.

However, to achieve the full controllability for systems under discrete linear time-invariant dynamics, the time cost is expensive, especially for dense networks \cite{ACSBabak}, where the relatively smaller number of driver nodes must broadcast the control signals to the entire network \cite{Liu2011,ACSBabak}. 
In this issue, a heuristic approach is introduced in \cite{ACSBabak} to identify the larger set of driver nodes compared with the MDNS in order to save the control time and computation cost. 
Specifically, Ravandi \textit{et al.} \cite{ACSBabak} first divide the nodes into four groups of equal size based on the in/out/total degree and they demonstrate that driver nodes in the MDNS are confined in the first-quartile of the in-degree distribution.
Staring from the statistical characteristics of MDNS regarding the  in-degree distribution, Ravandi \textit{et al.} \cite{ACSBabak} show that the heuristic approach offers a set of driver nodes with which a large proportion of nodes could be controlled at linear computational complexity.
For example, for empirical networks, they show 85\% nodes could be controlled at a velocity 378\% faster than the traditional framework to control all the nodes, with only 26\% more driver nodes compared with the number of nodes in the MDNS.

Under the classical linear time-invariant dynamics, the evolution of the system's states is determined by the current states, namely it is Markovian time dependent.
In this issue, Cao \textit{et al.} \cite{ACSCao} investigate the controllability of the discrete-time fractional-order linear dynamical networks \cite{Sergio18C}, where the system's non-Markovian time properties are incorporated with long-term memory. 
They present the trade-offs between the MDNS and the time to control based on the concept of actuation spectrum \cite{Sergio17SR}.  
Since the problem of determining the MDNS in this case is NP hard \cite{Sergio16C}, the authors offer a greedy approximation algorithm to calculate suboptimal solution of the actuation spectrum in polynomial time.
When the time-to-control is not larger than $\sim5\%$ of the network size, there is no significant difference in terms of the actuation spectrum for both artificial and empirical networked systems.
Interestingly, when the time-to-control is larger than the network size, the introduction of the fractional-order dynamics reduces the minimum number of driver nodes required to accomplish the network controllability.    
The change of MDNS under fractional-order dynamics ignites its possibility to dictate the dynamical characteristics over different networks.

\section{Nodes' role in control}
\begin{figure}
\centering
\includegraphics[width=0.82\textwidth]{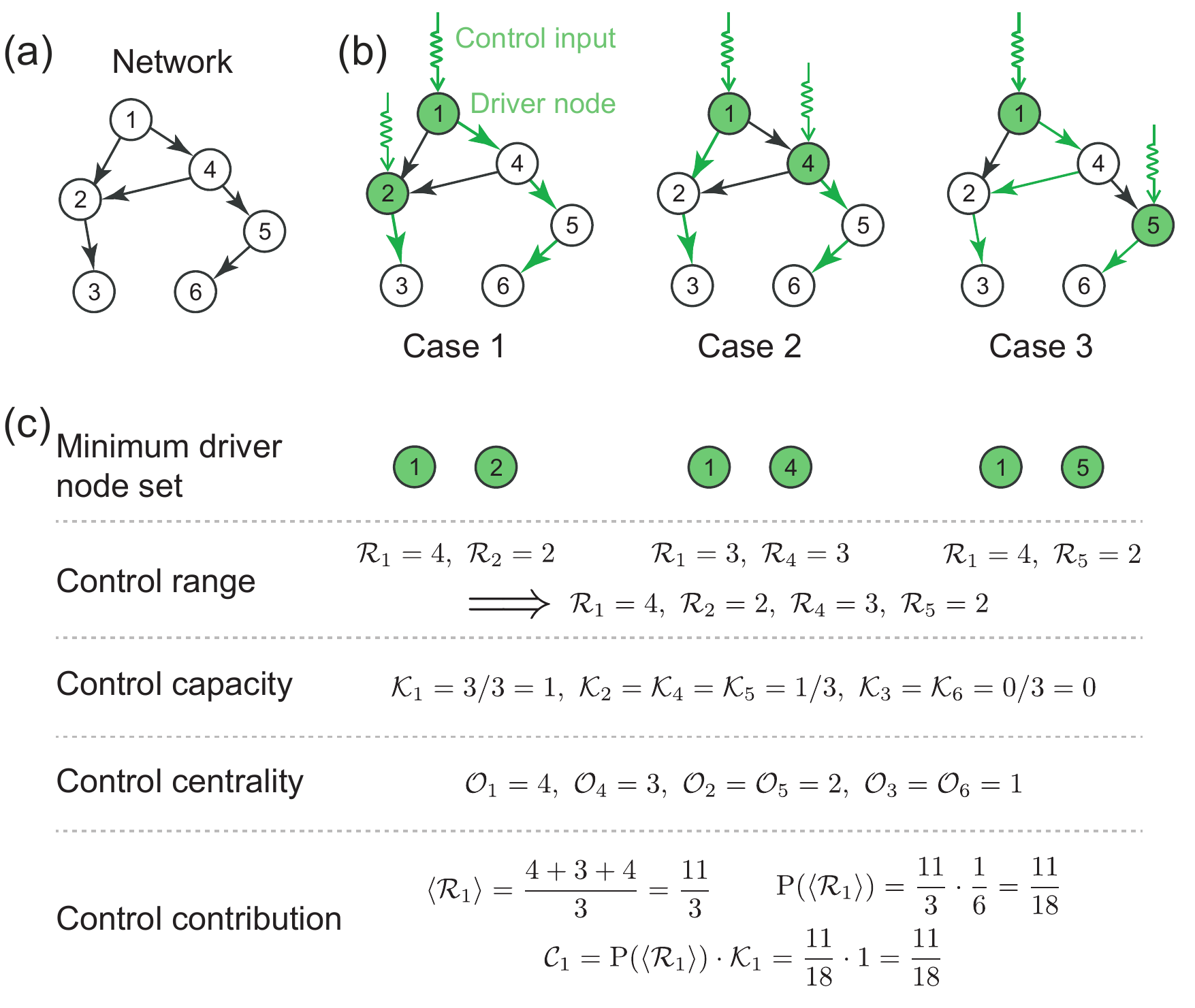}
\caption{
\textbf{Network controllability and nodes classification.}
(a) A directed network with $6$ nodes.
According to the method named maximum matching, we could find driver nodes for controlling complex networks \cite{Liu2011}.
Specifically, we first need to find the maximum set of directed links where they do not share the common start or end nodes, and then the end nodes of these links are matched.
The unmatched nodes are the driver nodes, for which we need to add independent control inputs (green jagged arrows) to fully control the network.
In (b), we show three cases of the maximum set of directed links (green links) according to the maximum matching, and matched (unmatched) nodes are in white (green).
For the three cases in (b), there are three minimum driver node sets (MDNSs) as shown in the first row of the panel (c).
The control range $\mathcal{R}_i$ is defined as the maximum size of the subnetwork that a driver node $i$ can control, and for each case, we list the control range of each driver node accordingly \cite{WangBingbo}.
For a network, a node's control range is chosen from the maximum value over all MDNSs.
And it quantifies the maximum ability node $i$ could exert to control the network together with other driver nodes.
The control capacity of each node $i$ (denoted by $\mathcal{K}_i$) -- the fraction of MDNS the node participates -- are presented in the third row in (c).
In other words, $\mathcal{K}_i$ indicates the probability for $i$ to be the driver node for a given MDNS.
When we only choose a single node $i$ to control with the external input, the maximum size of the subnetwork that the driver node can control is defined as the node's control centrality $\mathcal{O}_i$ \cite{Liu2012PlosOne}, and it is shown in the fourth row.
Here $\mathcal{O}_i$ quantifies the independent ability of node $i$ to control the whole network.
For the control contribution of a node, it is defined as the product of the node's control capacity and the average control range over different MDNSs divided by the total number of nodes \cite{Frank2019}.
In the last row of (c), we list the example for calculating the control contribution of the node $1$. 
}
 \label{Fig_nodes}
\end{figure}

Although the number of nodes in the MDNS keeps constant for a given network, the elements in different MDNSs may vary.
This opens the space for further explorations on nodes' uneven role in achieving the controllability \cite{Liu2012PlosOne,Jia2013,Ruths2014science}.
An intuitive definition to quantify the role of a driver node $i$ in control is the control capacity $\mathcal{K}_i$, which tells the fraction of MDNS in which $i$ participates \cite{JiaSR2013} (see Fig.~\ref{Fig_nodes}). 
Results show that even though $\mathcal{K}_i$ is independent of the node's out-degree, it decreases with its in-degree \cite{JiaSR2013}.
This echoes the results found in \cite{Liu2011} that driver nodes tend to avoid the highly connected nodes.
Indeed, when a node's in-degree increases, it is less likely for it to appear in an MDNS, which further leads to the decrease of $\mathcal{K}_i$.
Furthermore, by classifying the critical, redundant and intermittent nodes which correspond to $\mathcal{K}_i$ equal to $1$, $0$ and $c~(0<c<1)$ separately, two distinct control modes were observed \cite{Jia2013}.
One is the centralized control where only a small fraction of nodes are involved in MDNS, and another is distributed control where the majority of nodes could act as driver nodes.

Although $\mathcal{K}_i$ dictates the probability for a node to be involved in an MDNS, it does not tell how many nodes that $i$ is able to control once it is selected as a driver node. 
There are two ways to quantify this effect, one is control centrality \cite{Liu2012PlosOne} and another is control range \cite{WangBingbo}.
Control centrality $\mathcal{O}_i$ defines how many nodes that $i$ could control in the network when it is selected as the sole driver node (see Fig.~\ref{Fig_nodes}(c)) \cite{Liu2012PlosOne}.
That is to say, $\mathcal{O}_i$ quantifies the independent ability of node $i$ to control the whole network.
Note that here the node $i$ does not necessarily appear in any MDNS.
For the control range $\mathcal{R}_i$ of node $i$, it is defined by first calculating how many nodes (denoted by $N_i$) that $i$ controls when it is a node in some MDNSs \cite{WangBingbo}.
And $\mathcal{R}_i$ is chosen as the maximum value of $N_i$ over all possible MDNS (see Fig.~\ref{Fig_nodes}(c)).
For nodes with $\mathcal{K}_i=0$, meaning that $i$ never appears in any MDNS, $\mathcal{R}_i$ is normally defined as $1$. 
Intuitively, $\mathcal{R}_i$ quantifies the maximum ability node $i$ could exert to control the network together with other driver nodes.

Indeed, after having many independent variables to explore each node's importance, it is still lacking a comprehensive angle to determine how a node contributes to the network's controllability.
In this issue, Zhang \textit{et al.} \cite{Frank2019} propose the variable control contribution $\mathcal{C}_i$ of node $i$ (see Fig.~\ref{Fig_nodes}).
Generally, $\mathcal{C}_i$ incorporates two aspects: one is the control capacity $\mathcal{K}_i$, telling the probability for a node to serve as a driver node.
The other is the probability $\text{P}(\langle \mathcal{R}_i \rangle)$ for a random node to be controlled by the driver node $i$, which mathematically is the average size of the subnetwork that $i$ could control normalized by the network size, namely $\langle \mathcal{R}_i \rangle / N$.
And $\mathcal{C}_i = \text{P}(\langle \mathcal{R}_i \rangle) \mathcal{K}_i$, meaning the overall importance of node $i$ (see Fig.~\ref{Fig_nodes}(c)).

After proposing an optimization method to calculate the distribution $P(\mathcal{C})$ for both real and synthetic networks, Zhang \textit{et al.} \cite{Frank2019} find that the distribution is often diverse with no specific pattern, and $P(\mathcal{C})$ is not determined by the degree distribution.  
Moreover, they demonstrate that the two aspects of control capacity and control range do not share a uniform pattern, suggesting that $\mathcal{C}_i$ could indicate additional information.
While for the control range, it is shown that low-degree nodes tend to have a high control range \cite{WangBingbo}, which again echoes results shown in \cite{Liu2011} -- low-degree nodes are more likely to be driver nodes in heterogeneous networks and hence exhibit high control range.  
Importantly, the driver nodes selected according to their control contributions generate a larger controllable space than other selections according to the single control capacity or control range, which unambiguously shows the necessity for further research on this new variable \cite{Frank2019}.


\section{Link dynamics}

The equation (\ref{EquDynNode}) encodes the nodal dynamics and presents the way to control the states of systems' components.
Also, regarding the network, there is another indispensable ingredient -- the link connecting different nodes.
Indeed, links transmit control signals received from their start nodes to their end nodes, making network control feasible.
The dynamical process on the links is modelled as the so-called switchboard dynamics in \cite{Nepusz2012Edgedynamics}.
Specifically, every link is regarded as a switchboard like device that broadcasts the signal from the inbound links to the outbound ones with a linear operator.

The analysis on the link dynamics brings significant modifications to the understanding of controllability properties accumulated from nodal dynamics.
For example, the transcriptional regulatory networks are reported particularly easy to control under the framework of link dynamics.
And the existence of correction between in- and out-degrees reduces the MDNS for both Erd\H{o}s-R\'enyi random \cite{Erdos1960} and scale-free \cite{Barabasi1999a} networks, while the latter is easier to control at the same average nodal degree \cite{Nepusz2012Edgedynamics}.
Therefore, the feasible region of nodal or link dynamics is determined by the research question at hand, and link dynamics could be regarded as the nodal dynamics as well but with another network structure.

Nevertheless, for link dynamics, there are several different ways to understand how they affect each other or diverse scenarios of link weights \cite{Allesina2012,Kat15Science,BacterialNet17Xiao}.
Under switchboard dynamics \cite{Nepusz2012Edgedynamics}, the dynamics of an arbitrary link from the node $j$ to $i$ (\ie, $\dot{x}_{ij}(t)$) only depends on itself and the states of links pointing to $j$. 
In other words, it has no direct relation with the state of either node $i$ or $j$.
Here in this issue, Xiang and Chen \cite{ACSXiang} introduce the linear relation between states of nodes and links, namely $\dot{x}_{ij}(t) = \alpha \dot{x}_{j}(t) + \beta \dot{x}_{i}(t)$, where $\alpha$ and $ \beta$ are nonzero parameters.
And they investigate the minimal link controllability of directed networks with the MDNS consisting of a single node under the framework of nodal dynamics.

For three typical digraphs (alabastrum, snapback and cycle), the sufficient and necessary conditions are presented in terms of the function of $\alpha$ and $ \beta$ to ensure the both the nodal and link controllability, respectively \cite{ACSXiang}.
Moreover, the cycle is reported to be crucial for the link controllability, in which the number of nodes governs the power of $\alpha$ and $ \beta$.
Furthermore, Xiang and Chen \cite{ACSXiang} extend their results to signed digraphs and they demonstrate that the minimal link controllability of a signed cycle is determined by the number of links with negative weights. 
As the emergence of the explorations on various link dynamics, it opens the door to deepen our understanding of network controllability from the perspective of links -- the problems of interest when building the network from isolated components.

\section{Attractor control}

In practical applications, we do not always require to ergodically implement control from any initial state to any final state.
That is to say, there are normally a few number of final states that the system is supposed to operate over.
For example, in systems biology, a common goal is to provide concrete ways (\textit{e.g.}, drug targets) controlling a system from an initial (undesired or diseased) state to one of its attractors (healthy states).  
Under both discrete logic-based \cite{Reka03JTB,Zanudo2015aaa} and mechanistic \cite{CellMechanisticModel,PNASMechanisticModel} models, several effective control strategies have been reported \cite{attractorC1,Zanudo2017aaa}. 
But why the logic-based (Boolean) model -- a drastic simplification from complex biomolecular systems -- presents reliable predictions remain elusive.

In this issue, for the system regarding restriction switch of the cell cycle, Rozum and Albert \cite{ACSAlbert} study the relationship of controllability obtained over three different models (Boolean, Hill kinetics, and reaction-based models).
By examining the validity of the parameters derived from the mechanistic model on the control strategy offered by the stable motifs (self-sufficient positive circuits) of a Boolean counterpart model, 
they find that the control strategies obtained from Boolean and mechanistic models are in good agreement.
And it is no small part because of the parameter robustness -- a frequently observed feature of biomolecular systems -- of stable motifs \cite{Reka18PlosCB,Reka18PRE}.
Moreover, they show that the analysis of stable motifs helps us determine the effect of parameter uncertainty on the prediction of control.
Indeed, based on the system under study, how to appropriately model the system dynamics and further design control strategies require much more attention for implementing practical control \cite{Cornelius2013NatCommun,Wanglezhi16NC,Motter15PRX,ACSAlbert,Frank19EJP}.

\section{Control energy}
Beyond discussing the existence of admissible external control inputs to identify the full controllability of complex networks, we may need to estimate how much energy or cost it will take before implementing control in practice.
For a given pair of initial ($\X_0$) and final ($\X_\text{f}$) states, there are enormous ways to drive the system from $\X_0$ at time $t_0$ to $\X_\text{f}$ at time $t_\text{f}$ when the system is controllable.
Indeed, we could specify an arbitrary intermediate state in advance to make different trajectories.
Conventionally, the input energy \cite{OptimalBooLewis,Egerstedt2000CDC,Yan2012PRL,FPenergy2014,Yan2015a,Li2017Sci,chen2016,LaiYC2016RS,Li2017C,Klickstein2017,Duan2018} is defined as
\begin{equation*}
\label{InputEnergy}
E(\X_0,\X_\text{f}) =  \frac{1}{2}\int_{t_0}^{t_\text{f}}\U^\mathrm{T}(\tau)\U(\tau)\mathrm{d} \tau.
\end{equation*}
According to optimal control theory, there exists an optimal input $\U^*(t) = \B^\textrm{T}\textrm{e}^{\A ^\textrm{T} (t_\text{f} - t)} \W^{-1} \D $, with which the minimal input energy is 
\begin{equation}
\label{MInputEnergy}
E^*(\X_0,\X_\text{f}) =  \frac{1}{2} \D^\mathrm{T} \W ^{-1} \D. 
\end{equation}
Here $\W = \int_{t_0}^{t_\text{f}}\textrm{e}^{\A (\tau-t_0)}\B\B^\textrm{T}$ $\textrm{e}^{\A ^\textrm{T}(\tau-t_0)}\mathrm{d} \tau $ is the controllability Gramian matrix, 
and $\D = \X_\text{f} -  \textrm{e}^{\A (t_\text{f} - t_0)} \X_0$ is a vector pointing to the desired final state ($\X_\text{f}$, with control), from the natural final state that the system would reach without control.

\begin{figure}
\centering
\includegraphics[width=0.8\textwidth]{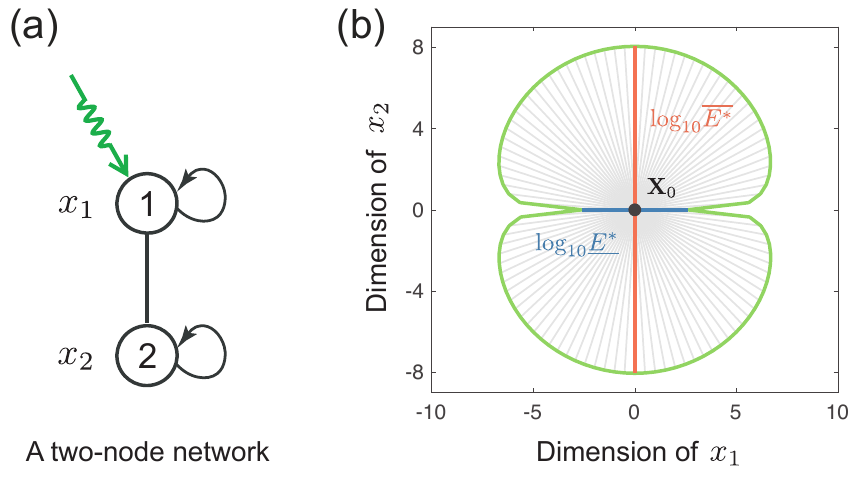}
\caption{
\textbf{Control energy for different final states.}
(a) For the left network, the full controllability is achieved by controlling the first node directly.
(b), For the initial state $\X_0=\zero$, we choose $100$ final states evenly satisfying $\| \X_\text{f} \|=1$.
For each $\X_\text{f}$, after calculating the minimal control energy $E^*(\X_0,\X_\text{f})$ according to the equation (\ref{MInputEnergy}), we draw a line from $\X_0$ to $\X_\text{f}$, where the length equals to $\text{log}_{10} E^*(\X_0,\X_\text{f})$.
Among all the $\X_\text{f}$, the maximum (minimum) energy indicated by $\overline{E^*}$ ($\underline{E^*}$), \ie, the upper (lower) bound of the minimal control energy is highlighted by red (blue) lines.
For the two dimensional system, we have $a_{12}=a_{21}=0.331$ with the strength of self-loops $-1.331$ for both nodes, and the time for directly controlling the first node is $10^{-2}$.
}
 \label{Fig_energy}
\end{figure}

For all possible desired states $\X_\text{f}$ at the same distance to $\X_0$ (namely $\delta = \| \X_\text{f}-\X_0\|$ is fixed), Yan \textit{et al.} \cite{Yan2012PRL} analyze the scaling law of the lower ($\underline{E^*}$) and upper ($\overline{E^*}$)  bounds of all $E^*(\X_0,\X_\text{f}) $.
To be specific, they theoretically show that, for $\X_0=\zero$ and $t_0=0$, $\underline{E^*}=1/\lambda_{\max}$, where $\lambda_{\max}$ is the maximum eigenvalue of $\W$. 
By virtue of the power of the trace (namely, the sum of the main diagonal elements of a square matrix, and it is equal to the sum of all eigenvalues accordingly) to approximate the maximum eigenvalue, the scaling behavior of $\underline{E^*}$ is presented in \cite{Yan2012PRL}.
However, for a random $\X_\text{f}$ at a distance $\delta$, the minimum energy is determined by $\overline{E^*}$ instead of $\underline{E^*}$ (see Fig.~\ref{Fig_energy}).
For the upper bound $\overline{E^*}=1/\lambda_{\min}$ where $\lambda_{\min}$ is the minimum eigenvalue of $\W$, the scaling law is theoretically derived by Duan \textit{et al.} recently \cite{Duan2018}.

%
%

Previous analyses on the optimal control energy are all based on the scenario where the system is fully controllable \cite{Yan2012PRL,FPenergy2014,Yan2015a,LaiYC2016RS,Klickstein2017,Duan2018}.
Indeed, it is natural to analyze the minimal energy from $\X_0$ to $\X_\text{f}$ after finding the MSDN to control a given network.
However, on the one hand, it is quite challenging to control the entire network, especially when it is heterogeneous \cite{Liu2011} where most nodes must be chosen to receive inputs directly.
On the other hand, for most systems \cite{CohenTargetCon03,ImmuTargetC06,NetTragetC06}, fully controllability is not necessarily to be ensured for a specific task. 
Indeed, the partial (output) controllability of complex networks has been studied under the concept of target control \cite{Gao2014}.
Results show that the full structural controllability expands the MDNS required for target control, for which the network heterogeneity also increases the control efficiency.
However, the related control energy regarding the target control remains unknown.

Here in this issue, Duan \textit{et al.} \cite{Duan2019ACS}
analyze the minimal control energy from $\X_0$ to $\X_\text{f}$, where only some components of $\X_\text{f}$ are controllable.
Under the framework of target control, the scaling behavior of both the lower and upper bounds of the optimal control energy in terms of control time is analytically presented.
They show that the energy cost could be saved by several orders of magnitude when the target control is invoked.
Moreover, at a given number of state components, results show that different requirements for controlling different nodes exhibit a significant difference at the energy cost.

%
\section{Control trajectory}
To control the system state from $\X_0$ to $\X_\text{f}$ with the minimal control energy, the optimal trajectory is
\begin{equation*}
\X(t) = \mathrm{e}^{\A (t - t_0)} \X_0+ \int_{t_0}^{t} \mathrm{e}^{\A (t - \tau)} \B 
\B^\textrm{T}\textrm{e}^{\A ^\textrm{T} (t_\text{f} - \tau)}
 \mathrm{d} \tau
  \W^{-1} [\X_\text{f} -  \textrm{e}^{\A (t_\text{f} - t_0)} \X_0].
\end{equation*}
To study the optimal trajectory, Sun and Motter \cite{Sun2013prl} propose the concept of nonlocality.
An initial state $\X_0$ is defined as strictly locally controllable: for a ball (with radius $\epsilon$) centered at $\X_0$,  if there is a concentric ball (with radius $\delta$, and $\delta \leq \epsilon$) such that any desired state $\X_\text{f}$ inside the 
ball with radius $\delta$ can be reached from $\X_0$ with a control trajectory totally inside the ball with radius $\epsilon$.
They found that the control trajectories are nonlocal generally, and numerical control fails for linear systems if $\W$ is ill-conditioned.
Nevertheless, the failure can be overcome by increasing the number of control inputs above the so-called numerical controllability transition \cite{Sun2013prl}. 
%
%

To capture the nonlocality of trajectories, two variables are required.
One is the length ($\mathcal{L}$), which captures how long the system state wanders before reaching the desired one \cite{Sun2013prl}.
For a given $\X_0$, the average length of the trajectories was calculated in \cite{Sun2013prl} over many different $\X_\text{f}$, which are chosen from the sphere centered at $\X_0$ with control distance $\delta$. 
Results show that the average $\mathcal{L}$ increases with $\delta$ when $\X_0=\zero$, while keeps as a constant for $\X_0 \neq \zero$ \cite{Sun2013prl}.
When $\delta$ is big enough, $\mathcal{L}$ follows the same scaling behavior for $\delta$ irrespective of whether $\X_0$ is at the origin or not \cite{Li2018Traj}.
Another variable is the radius ($\mathcal{R}$), which captures the maximum distance that the control trajectory deviates from the initial state among all of the system's intermediate states \cite{Li2018Traj}.
For the statistical behavior of control trajectories, $\mathcal{L}$ and $\mathcal{R}$ follow the similar scaling behavior in terms of $\delta$ \cite{Li2018Traj}.

\begin{figure}
\centering
\includegraphics[width=0.9\textwidth]{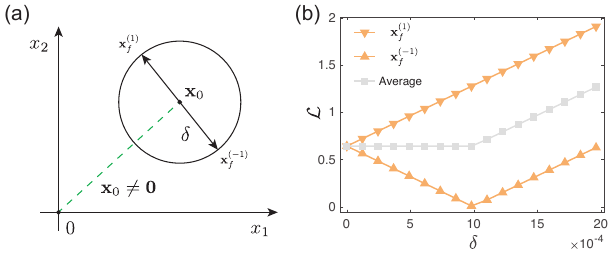}
\caption{
\textbf{Scaling behavior of the optimal control trajectory.}
(a) Illustration of a two dimensional system with the initial state $\X_0$, which is not at the origin.
For the two final states $\mathbf{x}_\text{f}^{(1)}$ and $\mathbf{x}_\text{f}^{(-1)}$, they both locate at the circle centered at $\X_0$
with radius $\delta$ but at opposite direction in terms of $\X_0$.
(b) We plot the length ($\mathcal{L}$) of the optimal control trajectory from $\X_0$ to $\mathbf{x}_\text{f}^{(1)}$ (lower triangle) and to $\mathbf{x}_\text{f}^{(-1)}$ (upper triangle) at different $\delta$.
The grey square is the average $\mathcal{L}$ for $\mathbf{x}_\text{f}^{(1)}$ and $\mathbf{x}_\text{f}^{(-1)}$.
Here we can see that the scaling behavior of $\mathcal{L}$ is different for $\mathbf{x}_\text{f}^{(1)}$ and $\mathbf{x}_\text{f}^{(-1)}$, while the average $\mathcal{L}$ keeps as a constant when $\delta<10^{-3}$.
Other parameters are the same as those in Fig.~\ref{Fig_energy}.
}
 \label{Fig_trajectory}
\end{figure}

Beyond the statistical behavior by averaging over different $\X_\text{f}$, the single behavior of $\mathcal{L}$ ($\mathcal{R}$)  for each specific $\X_\text{f}$ is worth enough attention as well, since they may be totally different \cite{Li2018Traj}.
For example, when $\X_0 \neq \zero$, the average $\mathcal{L}$ keeps as a constant for small $\delta$.
In this case, for a single $\X_\text{f}^{(1)}$ at one direction from $\X_0$, if $\mathcal{L}$ increases with $\delta$, then for another $\X_\text{f}^{(-1)}$ at the inverse direction (compared to the previous $\X_\text{f}^{(1)}$) from $\X_0$, $\mathcal{L}$ will decrease with $\delta$ (see Fig.~\ref{Fig_trajectory}).
Consequently, the different behavior of $\mathcal{L}$ for $\X_\text{f}^{(1)}$ and $\X_\text{f}^{(-1)}$ disappears after averaging them together, suggesting that the average cannot represent the behavior of every single element.
The distribution of $\mathcal{L}$ ($\mathcal{R}$) is presented in \cite{Li2018Traj} along different directions of final states at the same control distance.


\vspace{2cm}
\section{Discussions}

From detecting the controllability of networks to discussing the practical elements to implement control, we could tell that many findings around network dynamics and control are intimately linked to the underlying architecture of complex systems.
Sometimes the controllability may be lost by solely flipping the direction of a single link \cite{Liu2012PlosOne,Jia2013,Wang2012PRE}.
These factors further prompt us to pay additional attention back to the specific way by which natural, technological and social network structures are extracted. 

For systems where a physical or intuitive meaning of links between different components applies, the underlying networks could be built directly.
For example, the power-grid network is often built along the high-voltage power transmission lines connecting generators or substations \cite{Buldyrev2010,Brummitt2012}.
Likewise, brain networks could be constructed according to the number of white matter streamlines linking different regions \cite{Gu2016,Bassett18NP,TangReviewRMP}.
Ants interaction networks are often constructed in a convincing way where links represent the contact between the antenna of one ant and the body of another ant \cite{BlonderAnt2011}.
In a similar way, human face-to-face interaction networks are recently collected by the SocioPatterns collaboration (http://www.sociopatterns.org), where individuals wearing radio badges are monitored for face-to-face communications \cite{Cattuto2010,Isella2011,Fournet2014}. 

Sometimes, however, it is necessary to go further to infer the network structure from raw data collected along specific features of interest.
For instance, the interaction network of birds usually requires the high-speed machine version camera to record the real flocking behavior first, then reconstruct the three-dimensional trajectories of all birds with meticulous imaging techniques and algorithms, and finally build links based on birds' topological interaction range \cite{FlockingDataNP16,FlockingMethodPRE}.
Another example comes to the construction of microbial interaction networks in synthetic human gut microbiome communities, which is currently obtained by first performing experiments in a mouse model with some preselected bacterial species, and then measuring their relative abundance in the cecum and ileum \cite{BacterNet12Data}.  
Subsequently, ecological models are employed to infer the overall network among different species \cite{BacterNet2013PCB,RekaBacterialNet15,BacterialNet17Xiao}.

There are indeed experiments conducted over each pairwise community to infer the underlying networks of several species \cite{PairwiseBacteri18}, but challenges still remain for large systems as well as higher-order interactions for both microbial communities \cite{Jeff2016,ShouLV17} and social systems \cite{Causality14NatCom,HighOrderXu16SA}.   
On the whole, for self-organized systems like the flocking behavior of a huge number of self-propelled birds without central coordination \cite{SarmBeha1984,FlockingReviVicsek12,RenFlocking08}, or the massive and diverse microbes dwelling on almost every surface of our body \cite{MicrobeHuman,Kat15Science,Kevin2016Review,Kevin17Nature}, more efficient algorithms and methods are aspired to extrapolate the underlying skeleton precisely. 

Nevertheless, the availability of more useful data and precise networks is not the eventual quest.
Especially for problems emerged from interdisciplinary areas such as sociophysics, econophysics, biophysics and computational social science, upfront investigations and survey on requisite knowledge for problems (such as the specific meaning of the nodes, links and connectivity patterns) in mind play a key role at least equally compared with pursing embedded patterns or universal statistical laws \cite{FrankPhysicsToday}.
Indeed, pragmatic prerequisite investigations facilitate us to choose the appropriate network dynamics and even the parameters, which are all able to alter our final operation methods regarding systems control.

Furthermore, 
the unequal role and importance of different links/nodes in achieving control \cite{Liu2012PlosOne,Wang2012PRE,Jia2013,Ruths2014science,Zhoutong2019,Frank2019} reminds us that the implementation of control should not necessarily confine to external inputs injected directly on driver nodes or links \cite{Kalman63,Lin1970,Liu2011,Nepusz2012Edgedynamics,Wang2013,Liu2013Aut,Chen2014,Ruths2014science,Posfai2014NJP,Li2017Sci,StruCon,Liu2013PNAS,Cornelius2013NatCommun,Nepusz2012Edgedynamics}.
Rather, dynamical change of links and nodes provides another angle to understand and control some complex systems \cite{Holme2012,Holme2015review,NaokiBook,Naoki19SR,Christian19Entropy,Giacomo19arXiv,Li2017Sci}. 
For example, with the new components and connections added (active) and deleted (inactive) in adaptive ``smart'' infrastructures, research on controlling either links or nodes could provide smart strategies to better orchestrate the rhythm or time-scale of the network structure to achieve more efficient control at a lower cost.
Moreover, this could also provoke further explorations on defining and designing  distributed or centralized control regimes under which the network operates efficiently.

The temporality of real networks has been widely reported, which conceivably adds an additional dimension of time -- the dynamics of the network itself -- to capture dynamical systems \cite{Holme2012,Holme2015review,NaokiBook,Li2017Sci,Takayuki2019a}.
Indeed, links in metabolic networks correspond to brief chemical reactions \cite{Jeong2000},  and friendship links in social networks often indicate face-to-face or digital communications of short duration \cite{Cattuto2010,Isella2011,Fournet2014}.
Network temporality also exhibits its power on altering evolutionary outcomes of several dynamical processes like the diffusion or epidemic spreading \cite{Slowingdon12PRE,Slowingd13PRL,Ribeiro13SR,Causality14NatCom,Holme16PRE}, accessibility \cite{Accessibility13PRL} and the evolution of cooperative behavior \cite{Perc2006,Helbing09PNAS,Li2013a,Li2016arXiv}.
In addition, it is also shown that temporal networks could confer fundamental advantages in terms of control compared to their static counterparts, requiring less time to reach fully controllable, less control energy and also less length of optimal control trajectory \cite{Li2017Sci}.
Future investigations along this line may provide us practical ways to control diverse systems where either links or nodes are essentially time-varying in dynamical environments.

\section*{Acknowledgements}
We wish to thank G. Chen, F. Schweitzer, R. Albert, L. Wang, S. Pequito, B. Ravandi, F. Ansari, F. Mili, Y. Zhang,
J. C. Rozum, L. Xiang, G. Duan, T. Meng and Q. Cao
who contribute to the topical issue and provide constructive comments on this manuscript.
%
A.L. is supported by the International Human Frontier Science Program (HFSP) Postdoctoral Fellowship (Grant No. LT000696/2018-C) and Foster Lab at Oxford.
This paper has been accepted for publication at \textit{Advances in Complex Systems} in vol.~22, 2019.
We appreciate the patient help and guidance from Professor Frank Schweitzer.

%

\newpage

\linespread{1.1} 

\newpage

\newpage

\end{document}